# Stochastic resolution of identity to CC2 for large systems: Oscillator strength and ground state gradient calculations

Chongxiao Zhao,[1,2] Qi Ou,[3] Chenyang Li,[4,*] and Wenjie Dou[1,2]

[1]*Department of Chemistry, School of Science, Westlake University, Hangzhou, Zhejiang 310024, China*

[2]*Institute of Natural Sciences, Westlake Institute for Advanced Study, Hangzhou, Zhejiang 310024, China*

[3]*AI for Science Institute, Beijing 100080, China*

[4]*Key Laboratory of Theoretical and Computational Photochemistry, Ministry of Education, College of Chemistry, Beijing Normal University, Beijing 100875, China*

An implementation of stochastic resolution of identity (sRI) approximation to CC2 oscillator strengths as well as ground state analytical gradients is presented. The essential 4-index electron repulsion integrals (ERIs) are contracted with a set of stochastic orbitals on the basis of the RI technique and the orbital energy differences in the denominators are decoupled with the Laplace transform. These lead to a significant scaling reduction from $O(N^5)$ to $O(N^3)$ for oscillator strengths and gradients with the size of the basis set, $N$. The gradients need a large number of stochastic orbitals with $O(N^3)$, so we provide an additional $O(N^4)$ version with better accuracy and smaller prefactor by adopting sRI partially. Such steep computational acceleration of nearly two or one order of magnitude is very attractive for large systems. This work is an extension to our previous implementations of sRI-CC2 ground and excited state energies and shows the feasibility of introducing sRI to CC2 properties beyond energies.

## I. INTRODUCTION

During the past decades, the CC2 model, formulated as an approximation to the full CCSD[1], has been implemented for a vast range of electronic structure properties, including excited state energies[2–7], gradients[8-12], transition moments[13-17], frequency dependent properties[18,19] and *et al.*[20] Combined with some scaling reduction techniques[21-25], such as RI approximation[26-28], the expensive 4-index electron repulsion integrals in the CC2 model can be decomposed into lower-rank tensors, mostly scaling as $O(N^5)$ (with $N$ being a measure of the system size). This computational cost can relieve the bottleneck of disk space and computational time to some extent, but it's still prohibitively high for the calculations of large-scaled systems, which greatly hinders the further promotion and application of this model.

As for the CC2 oscillator strengths and analytical gradients, many reports focus on improving the efficiency in the past years. In 2012, Winter and Hättig[10] reported a scaled opposite-spin (SOS) CC2 implementation. All the expensive steps scaling as $O(N^5)$ are replaced by fourth order schemes. An overall $O(N^4)$ scaling was displayed for the transition moment and analytical gradient and it can lead to a significant reduction of costs. Some other reports also improved the performance[8-12,15-17], but most only ended up with a smaller prefactor.

Recently, a stochastic approach to RI approximation, abbreviated as sRI approximation, has been formulated to mitigate the high computational cost. Focusing on the flexible factorization of the 4-index ERIs, an additional set of stochastic orbitals is constructed to achieve further tensor hypercontractions. Such a sRI approximation has been successful implemented in various electronic structure models, such as DFT[29,30], MP2[31-33], GF2[34-38] and *et al.*[39,40] and has showed an impressive performance, especially in systems with hundreds or even thousands of electrons. Inspired by these successful implementations, we have introduced the sRI approach to calculate the CC2 ground state and excited state energies and achieved a scaling reduction to $O(N^3)$.[41,42] Therefore, it's promising to further extend the sRI approximation to other CC2 properties, such as the oscillator strength and the ground state analytical gradient in this manuscript.

Besides, another technique called Laplace transform is also adopted to decouple the indices in the CC2 algorithm. This approach was first developed to eliminate the energy denominators in MP2 theory.[43-45] Since many terms in *ab inito* methods are analogical to the MP2 energy, it has been rapidly popularized. With the Laplace transform, many efforts[46-53] have been paid to pursue for a lower scaling and some achieved an $O(N^4)$ scaling, including in the CC2 properties[10,11]. Based on the modules in some packages[54-56], we attempt to combine the strengths of both techniques and reduce the computational cost of sRI-CC2 oscillator strengths and ground state analytical gradients.

The manuscript is organized as follows: In Section II, the detailed implementations of sRI-CC2 oscillator strengths and analytical gradients are demonstrated. In Section III, the performance of sRI-CC2 approach for a variety of molecules is evaluated in contrast with the RI-CC2 results and experimental data, with emphasis on the accuracy and the scaling of CPU time. Finally, Section IV gives a conclusion.

## II. THEORY

The notations in Table I are used to represent the items in the main text. All the numbers in the last column are proportional to the system size.

TABLE I. Summary of notations in the following equations.

| item | function or indice | total number |
|---|---|---|
| AO Gaussian basis function | $\chi_\alpha(r_1), \chi_\beta(r_1), \chi_\gamma(r_1), \chi_\delta(r_1), \ldots$ | $N_{AO}$ |
| auxiliary basis function indice | $P, Q, R, S, \ldots$ | $N_{aux}$ |
| general sets of AO indice | $\alpha, \beta, \gamma, \delta, \ldots$ | $N_{ao}$ |
| general sets of MO indice | $p, q, r, s, \ldots$ | $N_{mo}$ |
| occupied MO indice | $i, j, k, l, \ldots$ | $N_{occ}$ |
| unoccupied (virtual) MO indice | $a, b, c, d, \ldots$ | $N_{vir}$ |

### A. CC2 oscillator strength

In the CC2 formulation, the Hamiltonian can be transformed with the single excitation cluster operator $T_1$ to simplify the expressions of CC2 equations

$$\widehat{H} = exp(-T_1) \, H \, exp(T_1) \tag{1}$$

$$T_1 = \sum_{\mu_1} t_{\mu_1} \tau_{\mu_1} \tag{2}$$

Here $\mu_1$ is an arbitrary single excited determinant. $t_{\mu_1}$ and $\tau_{\mu_1}$ are respectively the single excitation amplitudes and operators.

The CC2 ground state amplitudes $t_{\mu_1}$ can be obtained by iteratively solving the amplitude equations[1]:

$$\Omega_{\mu_1} = \langle \mu_1 | \widehat{H} + [\widehat{H}, T_2] | \text{HF} \rangle = 0 \tag{3}$$

$$\Omega_{\mu_2} = \langle \mu_2 | \widehat{H} + [F, T_2] | \text{HF} \rangle = 0 \tag{4}$$

where $\Omega_{\mu_1}$ and $\Omega_{\mu_2}$ are the single and double excitation vectors, $\langle\mu_1|$ and $\langle\mu_2|$ are the single and double excitation manifolds, and $|HF\rangle$ the Hartree-Fock reference state. $F$ is the Fock operator.

In CC response theory[57], the CC2 excitation amplitudes and energy $\omega$ can be solved as the eigenvectors and eigenvalues of the non-symmetric Jacobian matrix[3]

$$A_{\mu_i \nu_j} = \frac{\partial \Omega_{\mu_i}}{\partial t_{\nu_j}} = \begin{pmatrix} \langle\mu_1|[\hat{H}, \tau_{\nu_1}] + [[\hat{H}, \tau_{\nu_1}], T_2]|HF\rangle & \langle\mu_1|[\hat{H}, \tau_{\nu_2}]|HF\rangle \\ \langle\mu_2|[\hat{H}, \tau_{\nu_1}]|HF\rangle & \delta_{\mu_2 \nu_2} \varepsilon_{\mu_2} \end{pmatrix} \tag{5}$$

where $\varepsilon_{\mu_2}$ denotes the differences between orbital energy $\varepsilon_p$

$$\varepsilon_{\mu_2} = \varepsilon_{aibj} = \varepsilon_a - \varepsilon_i + \varepsilon_b - \varepsilon_j \tag{6}$$

Since the doubles-doubles block $A_{\mu_2 \nu_2}$ is diagonal, the effective Jacobian matrix can be derived as

$$A^{eff}_{\mu_1 \nu_1} = A_{\mu_1 \nu_1} - \frac{A_{\mu_1 \gamma_2} A_{\gamma_2 \nu_1}}{\varepsilon_{\gamma_2} - \omega} \tag{7}$$

In similar schemes, some singles amplitudes ($r_{\mu_1}$ and $l_{\mu_1}$ for right and left excitation amplitudes, $\bar{t}_{\mu_1}$ and $\bar{M}_{\mu_1}$ respectively for ground-state Lagrangian multipliers and transition moment Lagrangian multipliers) can be determined by solving the non-linear equations below. The explicit expressions for these equations were introduced by Koch, Hättig and *et al.*[3,4,13,15,58] systematically.

$$A^{eff}(\omega) r_1 = \omega r_1 \tag{8}$$

$$l_1 A^{eff}(\omega) = l_1 \omega \tag{9}$$

$$\bar{t}_1 A^{eff}(0) = -\eta^{eff}_1 \tag{10}$$

$$\bar{M}_1 (A^{eff}(-\omega) + \omega 1) = -\bar{m}^{eff}_1 \tag{11}$$

In our previous work[41,42], the sRI algorithms for the CC2 ground state and excitation amplitudes have been presented. Conventionally, throughout the CC2 algorithm, the 4-index ERIs as well as several transformed ones are constructed on the fly within the RI approximation. In order to further reduce the scaling, the sRI approximation is introduced to make more effective tensor hypercontractions. A brief introduction of RI and sRI approximations can be found in the Appendix.

$$(\alpha\beta|\gamma\delta) \approx \frac{1}{N_s} \sum_{\xi=1}^{N_s} R^{\xi}_{\alpha\beta} R^{\xi}_{\gamma\delta} \equiv \left\langle R^{\xi}_{\alpha\beta} R^{\xi}_{\gamma\delta} \right\rangle_{\xi} \tag{12}$$

For some double quantities and intermediates, the 4-index ERIs appear in the numerators and their denominators are orbital energy differences in Eq. (6). Such intermediates like MP2 energy are hard to decouple directly. Combined with the Laplace transform, the denominators of orbital energy differences can be disassembled into simple tensor multiplications of 2-rank tensors in Eq. (13).

$$\frac{1}{\epsilon_{aibj}} = \int_0^\infty e^{-\epsilon_{aibj}t}\, dt \approx \sum_z^{N_t} w_z e^{-\epsilon_{aibj}t_z} = \sum_z^{N_t} w_z (e^{-\epsilon_{ai}t_z} e^{-\epsilon_{bj}t_z}) \tag{13}$$

Here $w_z$ and $t_z$ are respectively the weights and the grid points of the numerical quadrature. $N_t$ is the number of these quadrature points and we select $N_t = 7$ for modest accuracy.

In the subsequently subsection, we will mainly focus on the differences after introducing sRI. The calculations of oscillator strengths need the transition moment Lagrangian multipliers $\bar{M}_i^a$ and all the expressions for Eq. (11) are listed in Table II. As for more details about other amplitudes in Eqs. (8)-(10), the work by Baudin *et al.*[17] is recommended since they summarized these serial CC2 implementations and corrected many typos and ambiguities in the former reports.

TABLE II. Explicit expressions for the transition moment lagrangian multiplier equation.

| term | $\bar{\sigma}_{ai} = \sum_{bj} \bar{M}_j^b A_{bj,ai}^{eff}(-\omega)$ | $\bar{m}_{ai}^{eff}$ |
|---|---|---|
| 0 | $\sum_b E_{ba}\bar{M}_i^b - \sum_j E_{ij}\bar{M}_j^a$ | $\sum_b \bar{E}_{ba}\bar{t}_i^b - \sum_j \bar{E}_{ij}\bar{t}_j^a$ |
| G | $+\sum_{cdk} \bar{M}_{ik}^{dc}(ck\hat{\vert}da)$ | $+\sum_{cdk} [F_{ik}^{dc}(ck\hat{\vert}da) + \bar{t}_{ik}^{dc}(ck\bar{\vert}da)]$ |
| H | $-\sum_{ckl} \bar{M}_{lk}^{ac}(ck\hat{\vert}il)$ | $-\sum_{ckl} [F_{lk}^{ac}(ck\hat{\vert}il) + \bar{t}_{lk}^{ac}(ck\bar{\vert}il)]$ |
| I | $+\sum_{ck} [2(kc\vert ia) - (ka\vert ic)]C_k^c$ | $+\sum_{ck} [2(kc\vert ia) - (ka\vert ic)](\bar{C}_k^c + 2r_k^c)$ |
| J | $+\sum_{ck} [2(ck\hat{\vert}ia) - (ca\hat{\vert}ik)]\bar{M}_k^c$ | $+\sum_{ck} [2(ck\bar{\vert}ia) - (ca\bar{\vert}ik)]\bar{t}_k^c$ |

$$\bar{M}_{ij}^{ab} = F_{ij}^{ab} + \frac{2(ia\breve{\vert}jb) - (ib\breve{\vert}ja) + \hat{P}_{ij}^{ab}[2\bar{M}_i^a \hat{F}_{jb} - \bar{M}_j^a \hat{F}_{ib}]}{\varepsilon_i - \varepsilon_a + \varepsilon_j - \varepsilon_b - \omega} \qquad F_{ij}^{ab} = \frac{2(\widetilde{ia\vert jb}) - (\widetilde{ib\vert ja}) + \hat{P}_{ij}^{ab}[2\bar{t}_i^a \bar{F}_{jb} - \bar{t}_j^a \bar{F}_{ib}]}{\varepsilon_i - \varepsilon_a + \varepsilon_j - \varepsilon_b - \omega}$$

$$C_i^a = \sum_{bj} \hat{t}_{ij}^{ab} \bar{M}_j^b \qquad \bar{C}_i^a = \sum_{bj} \hat{r}_{ij}^{ab} \bar{t}_j^b$$

$$E_{ij} = \hat{F}_{ij} + \sum_{cdk} \hat{t}_{jk}^{dc}(kc\vert id) \qquad \bar{E}_{ij} = \bar{F}_{ij} + \sum_{cdk} \hat{r}_{jk}^{dc}(kc\vert id)$$

$$E_{ba} = \hat{F}_{ba} - \sum_{ckl} \hat{t}_{lk}^{bc}(kc\vert la) \qquad \bar{E}_{ba} = \bar{F}_{ba} - \sum_{ckl} \hat{r}_{lk}^{bc}(kc\vert la)$$

With sRI, the expressions of the transformed 4-ERIs are rewritten

$$(\widetilde{ia|jb}) = -\hat{P}_{ij}^{ab} \sum_{ck} [\bar{t}_i^c r_k^c (ka|jb) + \bar{t}_k^a r_k^c (ic|jb)] = \langle \tilde{R}_{ia}^{\xi} R_{jb}^{\xi} + R_{ia}^{\xi} \tilde{R}_{jb}^{\xi} \rangle_{\xi} \quad (14)$$

$$\tilde{R}_{ia}^{\xi} = -\sum_{ck} (\bar{t}_i^c r_k^c R_{ia}^{\xi} + \bar{t}_k^a r_k^c R_{ic}^{\xi}) \quad (15)$$

$$(ia\check{|}jb) = \hat{P}_{ij}^{ab} \sum_{\alpha\beta\gamma\delta} (\bar{\Lambda}_{\alpha i}^p \Lambda_{\beta a}^h + \Lambda_{\alpha i}^p \bar{\Lambda}_{\beta a}^h) \Lambda_{\gamma j}^p \Lambda_{\delta b}^h (\alpha\beta|\gamma\delta) = \langle \check{R}_{ia}^{\xi} \hat{R}_{jb}^{\xi} + \hat{R}_{ia}^{\xi} \check{R}_{jb}^{\xi} \rangle_{\xi} \quad (16)$$

$$\hat{R}_{ia}^{\xi} = \Lambda_{\alpha i}^p \Lambda_{\beta a}^h R_{\alpha\beta}^{\xi} \qquad \check{R}_{ia}^{\xi} = (\bar{\Lambda}_{\alpha i}^p \Lambda_{\beta a}^h + \Lambda_{\alpha i}^p \bar{\Lambda}_{\beta a}^h) R_{\alpha\beta}^{\xi} \quad (17)\sim(18)$$

Now we can combine both sRI approximation and Laplace transform to the tensor multiplications in the Table II and below is a typical example.

$$\sum_{cdk} F_{ik}^{dc} (ck\hat{|}da) = \langle \sum_{cdk} F_{ik}^{dc}(\xi') \hat{R}_{ck}^{\xi} \hat{R}_{da}^{\xi} \rangle_{\xi\xi'}$$

$$= \langle \sum_{cdk} \hat{P}_{ik}^{dc} \frac{2\tilde{R}_{id}^{\xi'} R_{kc}^{\xi'} - \tilde{R}_{ic}^{\xi'} R_{kd}^{\xi'} + 2\bar{t}_i^d \bar{F}_{kc} - \bar{t}_i^c \bar{F}_{kd}]}{\varepsilon_i - \varepsilon_d + \varepsilon_k - \varepsilon_c - \omega} \hat{R}_{ck}^{\xi} \hat{R}_{da}^{\xi} \rangle_{\xi\xi'}$$

$$= -\langle \sum_z^{N_t} w_z \sum_{cdk} \hat{P}_{ik}^{dc} e^{-\omega t_z} (2N_{id}^{\tilde{R},\xi'} N_{kc}^{R,\xi'} - N_{ic}^{\tilde{R},\xi'} N_{kd}^{R,\xi'} + 2N_{id}^{\bar{t}} N_{kc}^{\bar{F}} - N_{ic}^{\bar{t}} N_{kd}^{\bar{F}}) \hat{R}_{ck}^{\xi} \hat{R}_{da}^{\xi} \rangle_{\xi\xi'} \quad (19)$$

$$N_{ia}^{R,\xi'}(t_z) = R_{ia}^{\xi'} e^{(\varepsilon_i-\varepsilon_a)t_z} \qquad N_{ia}^{\tilde{R},\xi'}(t_z) = \tilde{R}_{ia}^{\xi'} e^{(\varepsilon_i-\varepsilon_a)t_z}$$

$$N_{ia}^{\bar{t}}(t_z) = \bar{t}_i^a e^{(\varepsilon_i-\varepsilon_a)t_z} \qquad N_{ia}^{\bar{F}}(t_z) = \bar{F}_{ia} e^{(\varepsilon_i-\varepsilon_a)t_z} \quad (20)\sim(23)$$

In the RI scheme, the double quantity $F_{ik}^{dc}$ is constructed on the fly and the multiplication of $F_{ik}^{dc}$ and $(ck\hat{|}da)$ scales as $O(N^5)$ conventionally. However, now we subtly decouple $F_{ik}^{dc}$ into the combination of some 2-index tensor in Eqs. (20)~(23). It's obviously that the multiplications in the last row of Eq. (19) can be achieved gradually, with a scaling of $O(N^3)$. Notice the 4-index ERIs in the numerator of $F_{ik}^{dc}$ is independent of $(ck\hat{|}da)$, so another set of sRI matrix is employed, denoted $\xi'$.

Similarly, apart from $F_{ik}^{dc}$, all the 4-index tensors (double quantities and modified 4-index ERIs) can be decoupled into the tensor multiplications of some 2-index tensors in analogical steps. Then they can be easily contracted with other lower rank tensors such as the single amplitudes or intermediates $C_i^a$ and so on. This is just the key strategy. Overall, the scaling can be reduced from $O(N^5)$ to $O(N^3)$ to solve the transition moment multipliers.

With all the amplitudes, we can calculate the one-particle density matrices $D_{pq}^{\eta}$ and $D_{pq}^{\xi}$ [17]. Subsequently, the left and right transition moments can be formed as

$$T_{0m}^{vj} = \sum_{pq} [D_{pq}^{\eta}(r) + D_{pq}^{\xi}(\bar{M})] \hat{V}_{pq}^{j} \quad (24)$$

$$T_{m0}^{V^j} = \sum_{pq} D_{pq}^{\xi}(l)\hat{V}_{pq}^{j} \tag{25}$$

and then electric dipole transition strengths $S_{0m}^{V^jV^j}$ and oscillator strengths $f_{0m}$ in the length gauge respectively as

$$S_{0m}^{V^jV^j} = T_{0m}^{V^j}T_{m0}^{V^j} \tag{26}$$

$$f_{0m} = \frac{2}{3}\omega_m \sum_j S_{0m}^{V^jV^j} \tag{27}$$

Here $\hat{V}_{pq}^j$ is a Cartesian component of the $T_1$-transformed electric dipole integrals in the length gauge and j indicates the components of the x, y and z axes. The scaling for these steps are all lower than $O(N^3)$, so it's the final scaling.

## B. CC2 ground state gradient

The CC2 ground state gradient is solved by following a similar algorithm of CCSD analytical gradients[59]. The CC2 Lagrange function is defined as

$$L_{CC2} = E_{CC2} + \sum_{ai}\bar{t}_{ai}\Omega_{ai} + \frac{1}{4}\sum_{aibj}\bar{t}_{ij}^{ab}\Omega_{ij}^{ab} + \sum_{pq}\zeta_{pq}(F_{pq} - \delta_{pq}\varepsilon_p) + \sum_{pq}\omega_{pq}(S_{pq} - \delta_{pq}) \tag{28}$$

On the right side of Eq. (28), it contains the CC2 ground state energy $E_{CC2}$ and constraints for the ground state amplitudes in the first three terms. The fourth term suggests the use of Hartree-Fock orbitals and the last term indicates the orthonormality of spin orbitals. Multipliers $\bar{t}_{ai}$ and $\bar{t}_{ij}^{ab}$ can be solved by Eq. (10). $\zeta_{pq}$ and $\omega_{pq}$ are undetermined Lagrangian multipliers. $S_{pq}$ is the overlap matrix. It's convenient to rewrite the Lagrangian as

$$L_{CC2} = \sum_{pq}\bar{\gamma}_{pq}h_{pq} + \sum_{pqrs}\bar{\gamma}_{rs}^{pq}(pq|rs) + \sum_{pq}\zeta_{pq}(F_{pq} - \delta_{pq}\varepsilon_p) + \sum_{pq}\omega_{pq}(S_{pq} - \delta_{pq}) \tag{29}$$

Here $h_{pq}$ is the 1-electron integral. The $\bar{\gamma}_{ai}$ and $\bar{\gamma}_{aibj}$ are respectively the unrelaxed 1- and 2-body reduced density matrix (RDM) elements of CC2. The orbital response multipliers $\zeta_{pq}$ and $\omega_{pq}$ are determined from a variational condition in Eq. (30) and it's just the so-called z-vector equation.

$$\sum_{\mu} C_{\mu p}\frac{\partial L_{CC2}}{\partial C_{\mu q}} = 0 \tag{30}$$

The $C$ is the SCF molecular orbital coefficient matrix. All the blocks of densities and the explicit expressions for Eq. (30) are summarized in the Appendix.

The analytical gradient to an external parameter $x$ is written as

$$\frac{\partial L_{CC2}}{\partial x} = \sum_{pq}\gamma_{pq}h_{pq}^{[x]} + \sum_{pqrs}\gamma_{rs}^{pq}(pq|rs)^{[x]} + \sum_{pq}\omega_{pq}S_{pq}^{[x]} \tag{31}$$

The final 1- and 2-RDMs are calculated by

$$\gamma_{pq} \leftarrow \bar{\gamma}_{pq} \qquad\qquad \gamma_{ia} \leftarrow \zeta_{ai} \qquad (32)\sim(33)$$

$$\gamma_{rs}^{pq} \leftarrow \bar{\gamma}_{rs}^{pq} \qquad\qquad \gamma_{ak}^{ik} \leftarrow \zeta_{ai} \qquad (34)\sim(35)$$

Analogous to the algorithm for oscillator strengths, all the intermediates can be solved with sRI and show an $O(N^3)$ scaling. It's worth mentioned that the second term on the right side of Eq. (31) is a little different. The 4-rank integrals $(pq|rs)^{[x]}$ can be contracted into lower-rank ones.

$$(pq|rs)^{[x]} = \sum_{PQ}(pq|P)^{[x]}V_{PQ}^{-1}(Q|rs) + \sum_{PQ}(pq|P)V_{PQ}^{-1}(Q|rs)^{[x]} - \sum_{PQRS}(pq|R)V_{RP}^{-1}V_{PQ}^{[x]}V_{QS}^{-1}(S|rs) \quad (36)$$

The expression for the two-electron contribution to gradient is also written for both RI and sRI schemes and we need to calculate the 3- and 2-index two particle densities $\Delta_{pq}^P$ and $\gamma_{PQ}$.

$$\sum_{pqrs}\gamma_{rs}^{pq}(pq|rs)^{[x]} = 2\sum_{Ppq}\Delta_{pq}^P(pq|P)^{[x]} - \sum_{PQ}\gamma_{PQ}V_{PQ}^{[x]} \qquad (37)$$

$$\Delta_{pq}^P = \sum_{Qrs}\gamma_{rs}^{pq}B_{rs}^Q V_{PQ}^{-1/2} = \langle \sum_{rs}\gamma_{rs}^{pq}R_{rs}^\xi \sum_{Q}V_{PQ}^{-1/2}\theta_Q^\xi \rangle_\xi \qquad (38)$$

$$\gamma_{PQ} = \sum_{Rpq}\Delta_{pq}^Q B_{pq}^R V_{RP}^{-1/2} = \langle \sum_{pq}\Delta_{pq}^Q R_{pq}^\xi \sum_{R}V_{RP}^{-1/2}\theta_R^\xi \rangle_\xi \qquad (39)$$

The most time consuming step is the contraction of $\gamma_{rs}^{pq}$ and $B_{rs}^Q$. In our sRI scheme, different blocks of 4-rank matrix $\gamma_{rs}^{pq}$ in Eq. (38) can be separately contracted with $R_{rs}^\xi$ and there is no need to calculate the whole $\gamma_{rs}^{pq}$. We take the solution of one block as an example in Eq. (40). The scaling for each block is reduced to $O(N^3)$. Also, Eq. (39) scales as $O(N^3)$.

$$\sum_{rs}\gamma_{rs}^{pq}R_{rs}^\xi \leftarrow \sum_{bj}\hat{t}_{ij}^{ab}R_{bj}^\xi = \langle \sum_{bj}\frac{2R_{ai}^{\xi'}R_{bj}^{\xi'} - R_{bi}^{\xi'}R_{aj}^{\xi'}}{\varepsilon_i - \varepsilon_a + \varepsilon_j - \varepsilon_b}R_{bj}^\xi \rangle_{\xi'}$$
$$= -\langle \sum_z^{N_t}w_z \sum_{bj}\left(2N_{ai}^{R,\xi'}N_{bj}^{R,\xi'} - N_{bi}^{R,\xi'}N_{aj}^{R,\xi'}\right)R_{bj}^\xi \rangle_{\xi'} \qquad (40)$$

Now all the terms scaling as $O(N^5)$ can be reduced to $O(N^3)$ with the combination of sRI approximation and Laplace transform technique. However, tests with our $O(N^3)$ code show that a very large number of stochastic orbitals is needed to achieve proper accuracy for geometry optimization. Compared with the improving the accuracy with a larger $N_s$, another strategy is to adopt sRI partially in the exchange terms as Eq. (41) shows. To make it distinguishable, we name this process in Eq. (41) as 'partial sRI' and the one in Eq. (40) as 'complete sRI'.

$$\sum_{bj}\hat{t}_{ij}^{ab}B_{bj}^Q = \sum_{Pbj}\frac{2B_{ai}^P R_{bj}^P - \langle R_{bi}^{\xi'}R_{aj}^{\xi'}\rangle_{\xi'}}{\varepsilon_i - \varepsilon_a + \varepsilon_j - \varepsilon_b}B_{bj}^Q$$
$$= -\sum_z^{N_t}w_z \sum_{Pbj}\left(2N_{ai}^{B,P}N_{bj}^{B,P} - \langle N_{bi}^{R,\xi'}N_{aj}^{R,\xi'}\rangle_{\xi'}\right)B_{bj}^Q \qquad (41)$$

$$N_{ai}^{B,P}(t_z) = B_{ai}^P e^{(\varepsilon_i - \varepsilon_a)t_z} \tag{42}$$

The last row of Eq. (41) contains two terms. The former scales as $O(N^4)$ if we multiply $N_{bj}^{B,P}$ and $B_{bj}^Q$ first to cancel indices $b$ and $j$. The latter shows the same scaling by contracting $N_{bi}^{R,\xi'}$ and $B_{bj}^Q$ first. It's feasible to rewrite the whole CC2 codes with the partial sRI, including the former parts (ground state amplitudes and Lagrangian multipliers), so that there are less chances to introduce errors and standard deviations from sRI. In return, with the same number of stochastic orbitals, it shows proper accuracy at the sacrifice of a better scaling. Less stochastic orbitals is needed for such an $O(N^4)$ 'partial sRI' gradient calculations and it's fascinating when a commonly used $N_s$ could not satisfy the accuracy. In the next section, some data for both sRI-CC2 processes are provided for comparison.

## III. RESULTS AND DISCUSSION

In this section, our sRI-CC2 programs are applied to some typical molecules to test oscillator strengths and ground state gradients. The performance in accuracy and time consumption is discussed in comparison with the RI-CC2 approach. The analysis of the number of stochastic orbitals $N_s$ is also presented. If not stated otherwise, the calculations use the cc-pVDZ basis set. The estimation of the sRI stochastic error is performed by 10 independent runs with different random seeds and the final results are averaged. The error bars in the figures indicate the standard deviations and RI-CC2 data falling on the error bars prove the validity of sRI approximation. The energy criterion is set to 6 decimals and the residual criterion to 5 decimals. All the calculations are carried out in the high performance computing (HPC) center of Westlake University, utilizing an AMD EPYC 7502 (2.5GHz) node with 64 computational cores.

### A. CC2 oscillator strength

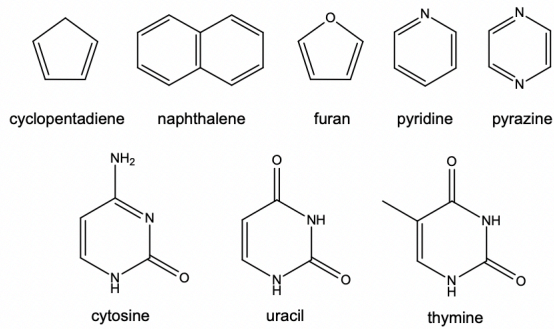

FIG. 1. Structural formulas of some medium-sized molecules.

Some medium-sized molecules in the Schreiber's test set[60] are first considered to assess the accuracy of our sRI-CC2 programs. In Table III, we report the oscillator strengths for the lowest electronic transitions of the molecules in Figure 1 with $N_s$ = 10000. We notice that the all the absolute (abs) errors are below 0.01. All the standard deviations (S.D.) cover the errors and are all within 0.02.

TABLE III. Oscillator strengths of some medium-sized molecules.

| molecule | RI | sRI | S.D. | abs error |
|---|---|---|---|---|
| cyclopentadiene | 0.1068 | 0.1069 | 0.0019 | 0.0001 |
| naphthalene | 0.1135 | 0.1121 | 0.0075 | 0.0014 |
| furan | 0.1797 | 0.1772 | 0.0060 | 0.0025 |
| pyridine | 0.0311 | 0.0300 | 0.0034 | 0.0011 |
| pyrazine | 0.0311 | 0.0300 | 0.0034 | 0.0011 |
| cytosine | 0.0541 | 0.0527 | 0.0062 | 0.0014 |
| uracil | 0.2076 | 0.2009 | 0.0117 | 0.0067 |
| thymine | 0.2155 | 0.2102 | 0.0158 | 0.0053 |

In Table IV, we apply our sRI-CC2 to a series of (all-$E$)-olefin chains with $N_s$ = 5000. It seems that both the oscillator strengths and standard deviations show a growing trend with the system size (Figure 2 shows an excellent goodness-of-fit), so we add the columns for these values per electron. The standard deviations are about 3% of the oscillator strengths while the absolute errors are below 1%. Averaged over the electron number, all these values basically remain in relatively stable ranges. This means that we don't need to increase the prefactor $N_s$ to achieve similar accuracy for larger systems.

TABLE IV. Oscillator strengths of a series of (all-$E$)-olefin chains.

| molecule | RI | sRI | S.D. | abs error | RI per e | sRI per e | S.D. per e | abs error per e |
|---|---|---|---|---|---|---|---|---|
| $C_2H_4$ | 0.4764 | 0.4811 | 0.0202 | 0.0047 | 0.02977 | 0.03007 | 0.00126 | 0.00029 |
| $C_4H_6$ | 0.8648 | 0.8682 | 0.0250 | 0.0034 | 0.02883 | 0.02894 | 0.00083 | 0.00011 |
| $C_6H_8$ | 1.3308 | 1.3409 | 0.0311 | 0.0101 | 0.03024 | 0.03047 | 0.00071 | 0.00023 |
| $C_8H_{10}$ | 1.7959 | 1.7995 | 0.0673 | 0.0036 | 0.03096 | 0.03103 | 0.00116 | 0.00006 |
| $C_{10}H_{12}$ | 2.2435 | 2.2435 | 0.0996 | 0.0187 | 0.03116 | 0.03090 | 0.00138 | 0.00026 |

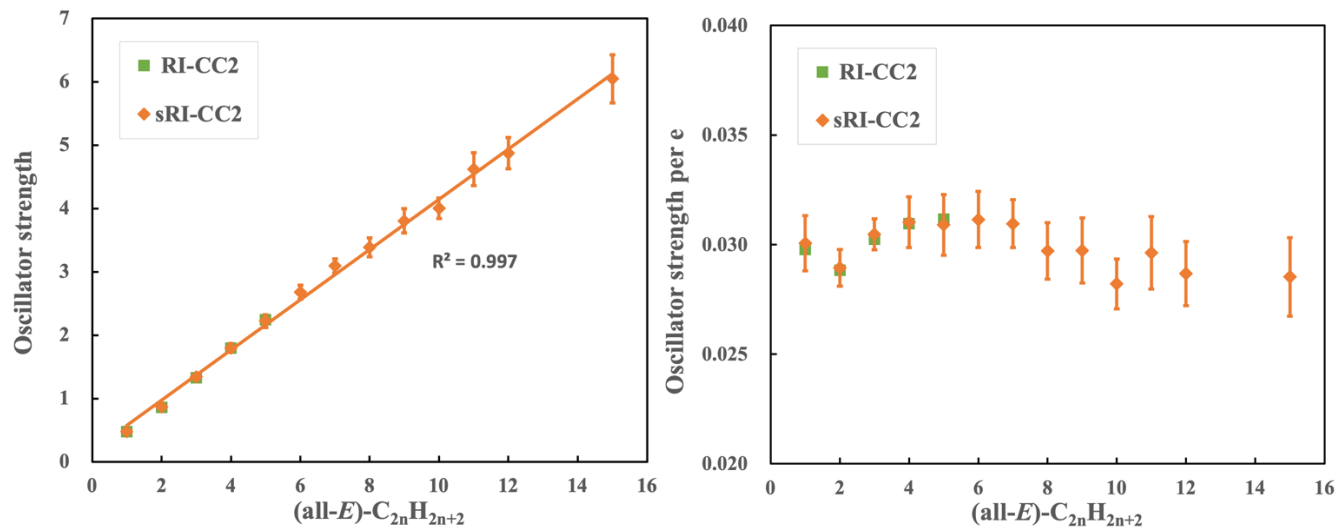

FIG. 2. Oscillator strengths of some olefin chains. The left subfigure shows the total values while the right one indicates the oscillator strengths per electron. The error bars indicate the standard deviations (S.D.).

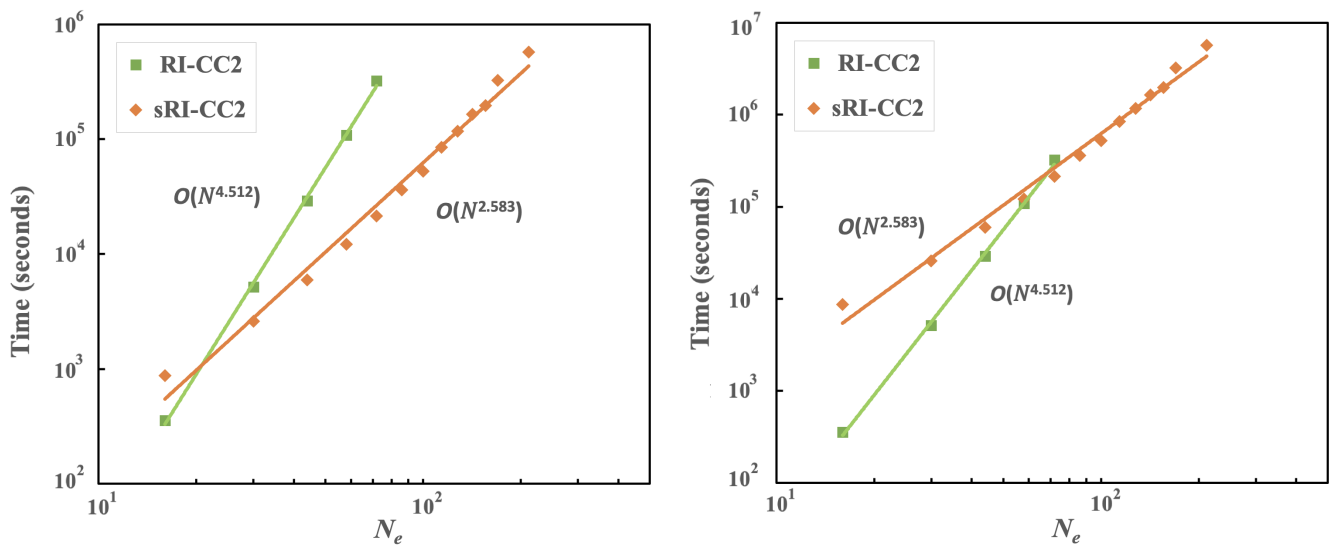

FIG. 3. CPU time as a function of the number of electrons. The left sRI-CC2 plot is from the averaged time while the right one is from the total time of 10 runs.

In Figure 3, we plot the CPU time of both RI-CC2 and sRI-CC2 as a function of the number of electrons. The crossover appears at about $N_e = 20$ (The joint is at $N_e = 70$ if we consider the total time). The RI-CC2 shows an observed scaling of $O(N^{4.51})$ and the sRI-CC2 $O(N^{2.58})$. Beyond the crucial $O(N^3)$ steps, other steps with lower ranks also influence the final scaling, resulting in an experimental cost better than the theoretical one. For systems larger than the crossover, our sRI-CC2 presents a superior performance than conventional RI-CC2, which is attractive for systems with hundreds or even thousands of electrons.

### B. CC2 ground state gradient

In order to measure the performance of our partial sRI-CC2 and complete sRI-CC2, some typical single-configuration dominated molecules in Ref. 61 are selected to calculate the equilibrium structures. We adopt $N_s = 100$ for partial sRI-CC2 and $N_s = 50000$ for complete sRI-CC2 and we will discuss the choices of $N_s$ in Sec. III C. The initial geometries are mainly taken from the reference[62]. For better comparison with the RI-CC2 results, all calculations for the bond lengths used the cc-pVTZ basis and a freezing-core approximation (freeze the previous noble gas shell on each atom) in Table V. Among our tests, we can notice that both absolute errors and the standard deviations are below 1% for partial sRI-CC2. It surely requires less stochastic orbitals to achieve better accuracy. In contrast, complete sRI shows larger absolute errors and fails to achieve similar accuracy even with $N_s = 50000$.

TABLE V. Bond length (in pm) of some typical molecules.

| molecule | bond | RI[8] | partial sRI | S.D. | abs error | complete sRI | S.D. | abs error | experiment[63-65] |
|---|---|---|---|---|---|---|---|---|---|
| $H_2$ | H-H | 73.73 | 73.78 | 0.139 | 0.05 | 73.83 | 0.152 | 0.10 | 74.15 |
| HF | H-F | 91.96 | 91.75 | 0.196 | 0.21 | 91.81 | 0.321 | 0.15 | 91.69 |
| $H_2O$ | O-H | 96.05 | 95.82 | 0.446 | 0.23 | 94.55 | 0.971 | 1.50 | 95.79 |
| $NH_3$ | N-H | 101.2 | 101.15 | 0.185 | 0.05 | 101.34 | 0.110 | 0.14 | 101.1 |
| BF | B-F | 127.4 | 126.85 | 0.330 | 0.55 | 132.21 | 6.488 | 4.81 | 126.3 |
| $H_2S$ | H-S | 133.6 | 133.35 | 0.543 | 0.25 | 136.94 | 3.704 | 3.34 | 134.6 |
| $F_2$ | F-F | 141.3 | 141.09 | 0.582 | 0.21 | 142.47 | 1.420 | 1.17 | 141.3 |

| | | | | | | | | |
|---|---|---|---|---|---|---|---|---|
| FOH | H-O | 96.98 | 96.43 | 0.687 | 0.55 | 98.75 | 4.400 | 1.77 | 96.86 |
| | O-F | 144.2 | 143.32 | 0.440 | 0.88 | 141.31 | 4.851 | 2.89 | 143.4 |

* The abs error is the absolute error between RI results from the Ref. 8 and our sRI results. Experimental data of BF is from Ref. 63 and H$_2$S from Ref. 64. Others are taken from Ref. 65.

In order to quantity the errors compared to the experimental data, we perform the statistical analysis with the following measures, respectively the mean error ($\bar{\Delta}$), the standard deviation ($\Delta_{std}$), the mean absolute error ($\bar{\Delta}_{abs}$) and the maximum error ($\Delta_{max}$). Here $\Delta_i$ is the error compared to the experimental data. The results among 9 bond lengths are listed in Table VI.

$$\bar{\Delta} = \frac{1}{n}\sum_{i=1}^{n} \Delta_i \tag{50}$$

$$\Delta_{std} = \sqrt{\frac{1}{n-1}\sum_{i=1}^{n}(\Delta_i - \bar{\Delta})^2} \tag{51}$$

$$\bar{\Delta}_{abs} = \frac{1}{n}\sum_{i=1}^{n} |\Delta_i| \tag{52}$$

$$\Delta_{max} = \max_{i}|\Delta_i| \tag{53}$$

From the table, we can observe that the $\bar{\Delta}$ and $\Delta_{max}$ are a little larger in the absolute value in our partial sRI-CC2, but in generally our partial sRI-CC2 exhibits a similar performance in contrast with RI-CC2. Complete sRI-CC2 shows several times each value. These analysis values could be further reduced if we employ larger $N_s$, but since the current $N_s$ = 50000 has been very large for a sRI implementation, we just take the partial sRI-CC2 with $N_s$ = 100 for measurement in the later part.

TABLE VI. Statistical analysis relative to experiment (in pm).

| method | $\bar{\Delta}$ | $\Delta_{std}$ | $\bar{\Delta}_{abs}$ | $\Delta_{max}$ |
|---|---|---|---|---|
| RI | 0.14 | 0.63 | 0.45 | 1.10 |
| partial-sRI | -0.18 | 0.53 | 0.33 | 1.25 |
| complete-sRI | 0.89 | 2.54 | 1.70 | 5.91 |

Table VII presents the partial sRI-CC2 gradients of some (all-$E$)-olefin chains with $N_s = 100$. Taking the RI results as the references, we can calculate the mean value and standard deviation of each gradient matrix and measure the maximum errors $\Delta_{max}$ and mean absolute errors $\bar{\Delta}_{abs}$ respectively. The $\Delta_{max}$ is below 0.003 hartree/bohr and the $\bar{\Delta}_{abs}$ below 0.001 hartree/bohr. In Figure 4, we can clearly observe that both $\Delta_{max}$ and $\bar{\Delta}_{abs}$ for the gradient or standard deviation do not increase obviously with the system size.

TABLE VII. Gradient comparison of some (all-$E$)-olefin chains (in hartree/bohr).

| molecule | gradient | | S.D. | |
| --- | --- | --- | --- | --- |
|  | $\Delta_{max}$ | $\bar{\Delta}_{abs}$ | $\Delta_{max}$ | $\bar{\Delta}_{abs}$ |
| $C_2H_4$ | 0.0011 | 0.0005 | 0.0031 | 0.0014 |
| $C_4H_6$ | 0.0018 | 0.0005 | 0.0042 | 0.0017 |
| $C_6H_8$ | 0.0012 | 0.0004 | 0.0041 | 0.0017 |
| $C_8H_{10}$ | 0.0025 | 0.0005 | 0.0036 | 0.0016 |
| $C_{10}H_{12}$ | 0.0018 | 0.0005 | 0.0040 | 0.0017 |

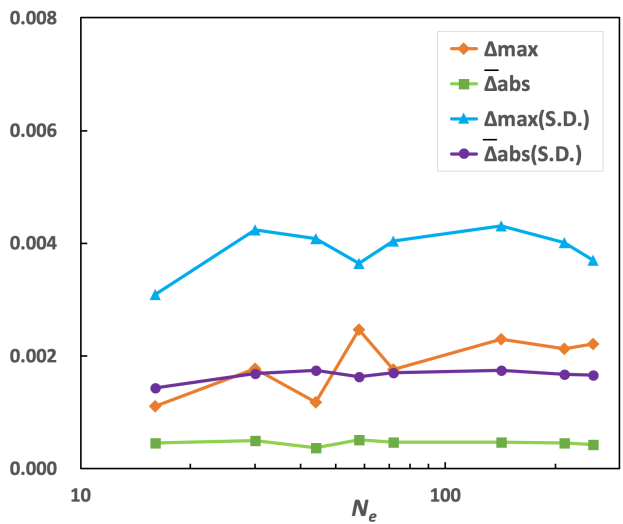

FIG. 4. Gradient assessment based on $\Delta_{max}$ and $\bar{\Delta}_{abs}$ for some (all-$E$)-olefin chains.

Figure 5 shows the time plots varying with the number of electrons for partial and complete sRI-CC2. The crossovers are predicted to happen at about $N_e = 200$ in both subfigures. Compared with the RI-CC2 plot, scaling as $O(N^{4.89})$, the experimental scaling is $O(N^{3.63})$ for partial sRI-CC2 and $O(N^{2.71})$ for complete sRI-CC2. Although the whole performance in the computational time couldn't compete with some packages (e.g., turbomole[66]), the scaling of our schemes is promising for large systems. Future progress mainly focuses on accelerating our current sRI program and simplifying the formulations.

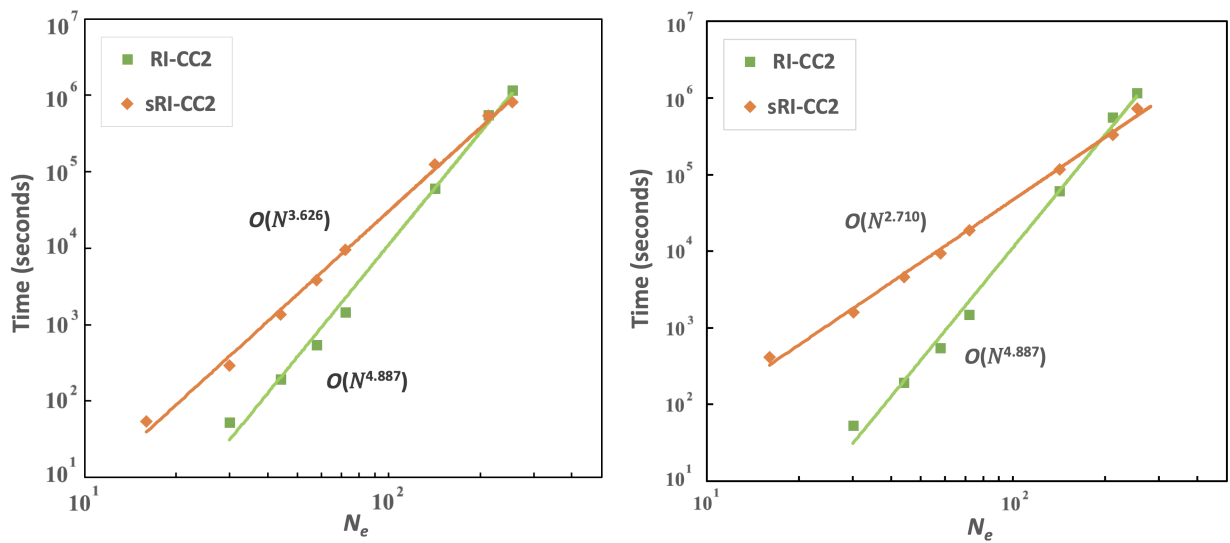

FIG. 5. CPU time as a function of the number of electrons. The left plot is from partial sRI-CC2 while the right one is from complete sRI-CC2.

### C. Assessment of the number of stochastic orbitals

In Figure 6, we take the water molecule as an example and assess the influence of the stochastic orbital number $N_s$ on the final oscillator strength. We can notice that the error bars decrease with stochastic orbitals and the sRI values gradually approaches the RI line. However, larger $N_s$ also enlarges the prefactor of the scaling and increases the computational cost. So, an effective sRI approach requires a compromise between accuracy and cost. Our strategy is to balance the two through a modest $N_s$. Our previous calculations show that $N_s = 800$ for the ground state energy and $N_s = 8000$ for the excitation energy are suitable for modest systems we test. Since these sRI-CC2 programs are continuous, we choose a larger $N_s = 10000$ for the oscillator strength calculations. The former programs to determine $t_{\mu_1}$, $\bar{t}_{\mu_1}$, $r_{\mu_1}$ and $l_{\mu_1}$ also employ $N_s = 10000$ so that we can shrink the influence of errors and standard deviations from the previous steps.

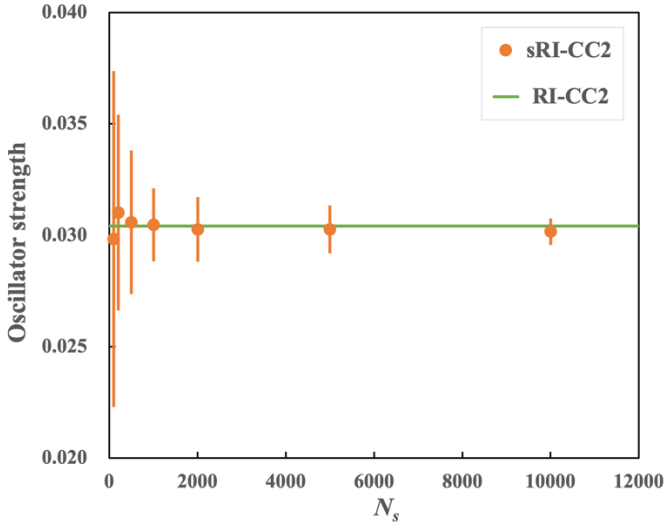

FIG. 6. Oscillator strengths for $H_2O$ under different number of stochastic orbitals.

Similarly, we evaluate the gradient of water molecule under different $N_s$ in Figure 7. The left subfigure is from partial sRI-CC2 and the right one from complete sRI-CC2. We can observe that all the plots decrease rapidly at first, and then tend to converge. For partial sRI-CC2, at about $N_s = 100$, both $\Delta_{max}$ and $\bar{\Delta}_{abs}$ for gradient start to go smoothly. For complete sRI-CC2, this appears at about $N_s = 50000$. Besides, even with a large number of stochastic orbitals, the error from complete sRI-CC2 is still relatively large and doesn't achieve the same order of partial sRI-CC2. Perhaps for large systems whose initial coordinates are far from equilibrium and for cases that don't require high accuracy, our complete sRI-CC2 can be adopted for its better performance in the scaling.

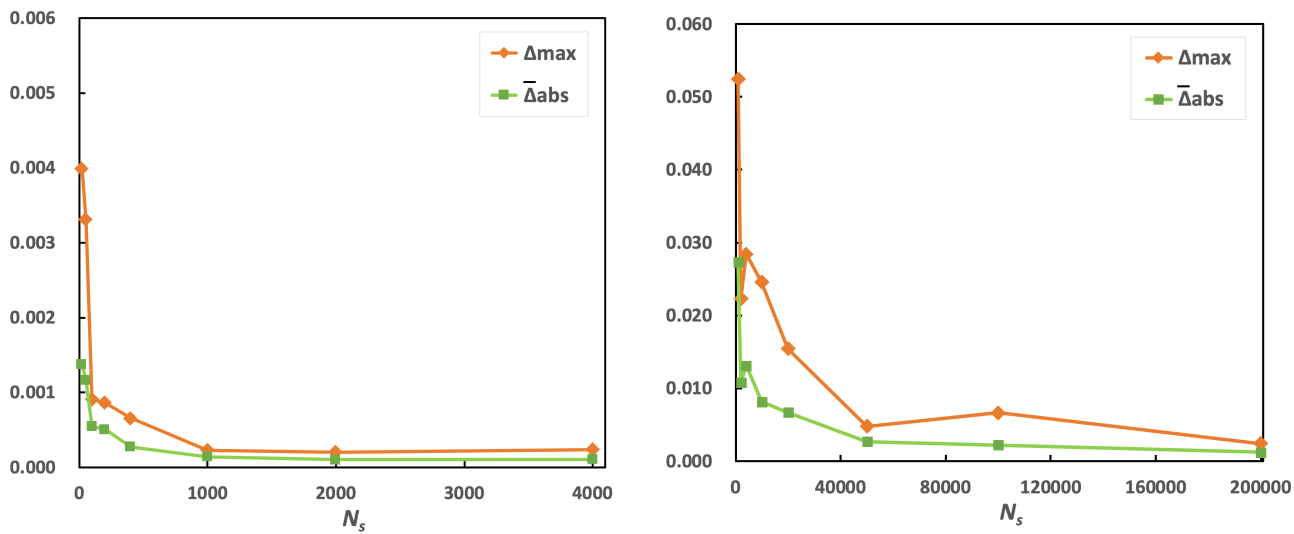

FIG. 7. Gradient assessment for $H_2O$ under different number of stochastic orbitals. The left subfigure is from partial sRI-CC2 and the right one from complete sRI-CC2.

## IV. CONCLUSIONS

We extend our sRI programs to CC2 response properties, e.g., oscillator strengths and ground state analytical gradients, and reduce the scaling to $O(N^3)$. Combined with the stochastic orbitals and Laplace transform, the sRI-CC2 oscillator strengths show a computational cost competing with that of HF and DFT and acceptable accuracy with a modest $N_s$. The complete sRI-CC2 analytical gradients scaling as $O(N^3)$ achieve a speedup of two orders of magnitude and we provide an $O(N^4)$ version with better accuracy by adopting sRI partially. Further efforts will be paid to simplify the formulations and reduce $N_s$, with the purpose of higher accuracy and less costs. Nonetheless, the current scaling reduction is promising for some large systems with hundreds of electrons. Our future interests may lie in other CC2 properties, such as excited state analytical gradients, derivative couplings, and the combination of sRI-CC2 with other theories.

## AUTHOR DECLARATIONS

### Conflict of Interest

The authors have no conflicts to disclose.

### ACKNOWLEDGEMENTS

We acknowledge the startup funding and high-performance computing (HPC) service from Westlake University. W.D. thanks the funding from National Natural Science Foundation of China (No. 22361142829) and Zhejiang

Provincial Natural Science Foundation (No. XHD24B0301). C. L. acknowledges the financial support from the National Natural Science Foundation of China (Nos. 22233001, 22103005) and the Fundamental Research Funds for the Central Universities. We are grateful for support and useful discussions from Joonho Lee. C. Z. thanks Yang Guo for helpful suggestions.

## APPENDIX A: INTRODUCTION OF RI AND SRI

In the RI approximation, the 4-index ERIs are contracted by introducing the so-called auxiliary basis $\{P\}$ to construct the 3-index and 2-index ERIs

$$(\alpha\beta|\gamma\delta) \approx \sum_{PR} (\alpha\beta|P)[V^{-1}]_{PR}(R|\gamma\delta) \tag{A1}$$

where the 4-, 3- and 2-index ERIs are defined as

$$(\alpha\beta|\gamma\delta) = \iint dr_1\, dr_2 \frac{\chi_\alpha(r_1)\chi_\beta(r_1)\chi_\gamma(r_2)\chi_\delta(r_2)}{r_{12}} \tag{A2}$$

$$(\alpha\beta|P) = \iint dr_1\, dr_2 \frac{\chi_\alpha(r_1)\,\chi_\beta(r_1)\,\chi_P(r_2)}{r_{12}} \tag{A3}$$

$$V_{PQ} = (P|Q) = \iint dr_1\, dr_2 \frac{\chi_P(r_1)\,\chi_Q(r_2)}{r_{12}} \tag{A4}$$

The common part can be defined as

$$B_{\alpha\beta}^Q \equiv \sum_P (\alpha\beta|P)\, V_{PQ}^{-1/2} \tag{A5}$$

Then the RI approximation can be rewritten as

$$(\alpha\beta|\gamma\delta) \approx \sum_Q B_{\alpha\beta}^Q B_{\gamma\delta}^Q \tag{A6}$$

The stochastic realization of the RI approximation introduces another set of stochastic orbitals $\{\theta^\xi\}$, $\xi = 1, 2, \ldots, N_s$. Here the $N_s$ is the number of these stochastic orbitals and is an empirical constant. All these stochastic orbitals are column arrays of length $N_{aux}$ and the elements are randomly generated to be 1 or -1 (i.e., $\theta_A^\xi = \pm 1$). So, these stochastic orbitals satisfy the following properties

$$\langle \theta \otimes \theta^T \rangle_\xi = \frac{1}{N_s} \sum_{\xi=1}^{N_s} \theta^\xi \otimes (\theta^\xi)^T$$

$$= \begin{pmatrix} \langle \theta_1\theta_1\rangle_\xi & \langle \theta_2\theta_1\rangle_\xi & \cdots & \langle \theta_1\theta_{N_{aux}}\rangle_\xi \\ \langle \theta_2\theta_1\rangle_\xi & \langle \theta_2\theta_2\rangle_\xi & & \langle \theta_2\theta_{N_{aux}}\rangle_\xi \\ \vdots & & \ddots & \vdots \\ \langle \theta_{N_{aux}}\theta_1\rangle_\xi & \langle \theta_{N_{aux}}\theta_2\rangle_\xi & \cdots & \langle \theta_{N_{aux}}\theta_{N_{aux}}\rangle_\xi \end{pmatrix} \approx I \tag{A7}$$

Since $\theta_A^\xi$ (and $\theta_B^\xi$) is a random choice of $\pm 1$, the diagonal matrix element denoted by $\langle \theta_A \theta_A \rangle_\xi$ always equals 1; the off-diagonal element denoted by $\langle \theta_A \theta_B \rangle_\xi$, however, converges to 0 with $N_s$ increases.

With the introduction of the sRI identity matrix in Eq. (A7), the original 4-index ERIs are transformed into a similar expression of the one with RI approximation:

$$(\alpha\beta|\gamma\delta) = \sum_{QS}\sum_{PR}(\alpha\beta|P)\,V_{PQ}^{-1/2}(\langle\theta\otimes\theta^T\rangle_\xi)_{QS}\,V_{SR}^{-1/2}(R|\gamma\delta)$$

$$= \left\langle \sum_P\left[(\alpha\beta|P)\sum_Q(V_{PQ}^{-1/2}\theta_Q)\right]\sum_R\left[(R|\gamma\delta)\sum_S(V_{SR}^{-1/2}\theta_S^T)\right]\right\rangle_\xi \qquad (A8)$$

Similar to the RI approximation, the common part is defined as

$$R_{\alpha\beta}^\xi = \sum_P\left[(\alpha\beta|P)\sum_Q(V_{PQ}^{-1/2}\theta_Q)\right] \qquad (A9)$$

And the 4-index ERIs are rewritten as

$$(\alpha\beta|\gamma\delta) \approx \frac{1}{N_s}\sum_{\xi=1}^{N_s} R_{\alpha\beta}^\xi R_{\gamma\delta}^\xi \equiv \left\langle R_{\alpha\beta}^\xi R_{\gamma\delta}^\xi \right\rangle_\xi \qquad (A10)$$

These integrals above are presented with atomic orbitals (AO) and can be easily transformed to molecular orbitals (MO) with the HF SCF coefficient. Since the number of stochastic orbitals, $N_s$, is independent from the system size, it only adds a prefactor to the final scaling and the overall calculation of 4-index ERIs with sRI technique approximately costs $O(N^4)$. Combined with the Laplace transform, this scaling can be further reduced in the CC2 algorithm and is discussed in Sec. II.

## APPENDIX B: DETAILED INTERMEDIATES

The explicit expressions for the one-particle density matrices and the corresponding double quantities are summarized in Table VIII.

TABLE VIII. Explicit expressions for the one-particle density matrices.

| block | $D_{pq}^\eta(r)$ | $D_{pq}^\xi(l)$ | $D_{pq}^\xi(\bar{M})$ |
|---|---|---|---|
| $D_{ij}$ | $-\sum_a \bar{t}_j^a r_i^a - \sum_{abk}\bar{t}_{jk}^{ab}r_{ik}^{ab}$ | $-\sum_{abk} l_{jk}^{ab} t_{ik}^{ab}$ | $-\sum_{abk}\bar{M}_{jk}^{ab}t_{ik}^{ab}$ |
| $D_{ia}$ | $2r_i^a + \sum_{ck}\bar{t}_k^c \hat{r}_{ik}^{ac} - \sum_b(\sum_{kjc}\bar{t}_{kj}^{bc}t_{kj}^{ac})r_i^b - \sum_j(\sum_{cbk}\bar{t}_{jk}^{cb}t_{ik}^{cb})r_j^a$ | $\sum_{ck} l_k^c \hat{t}_{ik}^{ac}$ | $\sum_{ck}\bar{M}_k^c \hat{t}_{ik}^{ac}$ |
| $D_{ai}$ | 0 | $l_i^a$ | $\bar{M}_i^a$ |

| $D_{ab}$ | $\sum_i \bar{t}_i^a r_i^b - \sum_{ijc} \bar{t}_{ij}^{ac} r_{ij}^{bc}$ | $\sum_{ijc} l_{ij}^{ac} t_{ij}^{bc}$ | $\sum_{ijc} \bar{M}_{ij}^{ac} t_{ij}^{bc}$ |
|---|---|---|---|

$$t_{ij}^{ab} = \frac{(ai\hat{|}bj)}{\varepsilon_i - \varepsilon_a + \varepsilon_j - \varepsilon_b} \qquad \hat{t}_{ij}^{ab} = 2t_{ij}^{ab} - t_{ij}^{ba}$$

$$r_{ij}^{ab} = \frac{(ai\bar{|}bj)}{\varepsilon_i - \varepsilon_a + \varepsilon_j - \varepsilon_b + \omega} \qquad \hat{r}_{ij}^{ab} = 2r_{ij}^{ab} - r_{ij}^{ba}$$

$$l_{ij}^{ab} = \frac{2(ia\breve{|}jb) - (ib\breve{|}ja) + P_{ij}^{ab}[2l_i^a \hat{F}_{jb} - l_j^a \hat{F}_{ib}]}{\varepsilon_i - \varepsilon_a + \varepsilon_j - \varepsilon_b + \omega}$$

The explicit expressions for z-vector equation and the densities in specific blocks for gradients calculations are summarized in Table IX.

TABLE IX. Explicit expressions for the orbital response term $\zeta_{pq}$ equation and the densities in spin orbitals.

$$0 = \sum_\mu C_{\mu a} \frac{\partial L}{\partial C_{\mu i}} = \sum_q h_{aq} \bar{r}_{qi} + \sum_p h_{pa} \bar{r}_{ip} + \sum_{qrs} v_{rs}^{aq} \bar{r}_{rs}^{iq} + \sum_{prs} v_{rs}^{pa} \bar{r}_{rs}^{pi} + \sum_{pqs} v_{as}^{pq} \bar{r}_{is}^{pq} + \sum_{pqr} v_{ra}^{pq} \bar{r}_{ri}^{pq} + \sum_b \zeta_{bi} F_{ba} + \sum_{bj} \zeta_{bj}(v_{ji}^{ba} + v_{ja}^{bi}) + 2\omega_{ia}$$

$$0 = \sum_\mu C_{\mu i} \frac{\partial L}{\partial C_{\mu a}} = \sum_q h_{iq} \bar{r}_{qa} + \sum_p h_{pi} \bar{r}_{ap} + \sum_{qrs} v_{rs}^{iq} \bar{r}_{rs}^{aq} + \sum_{prs} v_{rs}^{pi} \bar{r}_{rs}^{pa} + \sum_{pqs} v_{is}^{pq} \bar{r}_{as}^{pq} + \sum_{pqr} v_{ri}^{pq} \bar{r}_{ra}^{pq} + \sum_j \zeta_{aj} F_{ij} + 2\omega_{ia}$$

$$0 = \sum_\mu C_{\mu i} \frac{\partial L}{\partial C_{\mu j}} = \sum_q h_{iq} \bar{r}_{qj} + \sum_p h_{pi} \bar{r}_{jp} + \sum_{qrs} v_{rs}^{iq} \bar{r}_{rs}^{jq} + \sum_{prs} v_{rs}^{pi} \bar{r}_{rs}^{pj} + \sum_{pqs} v_{is}^{pq} \bar{r}_{js}^{pq} + \sum_{pqr} v_{ri}^{pq} \bar{r}_{rj}^{pq} + \sum_{ak} \zeta_{ak}(v_{kj}^{ai} + v_{ki}^{aj}) + 2\omega_{ij}$$

$$0 = \sum_\mu C_{\mu a} \frac{\partial L}{\partial C_{\mu b}} = \sum_q h_{aq} \bar{r}_{qb} + \sum_p h_{pa} \bar{r}_{bp} + \sum_{qrs} v_{rs}^{aq} \bar{r}_{rs}^{bq} + \sum_{prs} v_{rs}^{pa} \bar{r}_{rs}^{pb} + \sum_{pqs} v_{as}^{pq} \bar{r}_{bs}^{pq} + \sum_{pqr} v_{ra}^{pq} \bar{r}_{rb}^{pq} + 2\omega_{ab}$$

| | |
|---|---|
| $\bar{\gamma}_{ij} = \delta_{pq} - \sum_a \bar{t}_i^a t_j^a - \frac{1}{2}\sum_{abk} \bar{t}_{ik}^{ab} t_{jk}^{ab}$ | $\bar{\gamma}_{ia} = \bar{t}_i^a$ |
| $\bar{\gamma}_{ai} = t_i^a + \sum_{bj} \bar{t}_j^b(t_{ij}^{ab} - t_i^b t_j^a)$ | $\bar{\gamma}_{ab} = \sum_i \bar{t}_i^b t_i^a + \frac{1}{2}\sum_{ijc} \bar{t}_{ij}^{cb} t_{ij}^{ca}$ |

| | | |
|---|---|---|
| $\bar{\gamma}_{ij}^{ij} \leftarrow \frac{1}{2}$ | $\bar{\gamma}_{kl}^{cd} \leftarrow \sum_{ai} \bar{t}_i^a t_l^d t_{ik}^{ac}$ | $\bar{\gamma}_{ka}^{cd} \leftarrow \frac{1}{2}\sum_i \bar{t}_i^a t_{ki}^{cd}$ |
| $\bar{\gamma}_{ik}^{ak} \leftarrow t_i^a$ | $\bar{\gamma}_{kl}^{cl} \leftarrow \sum_{ai} \bar{t}_i^a t_{ik}^{ac}$ | $\bar{\gamma}_{kl}^{ci} \leftarrow -\sum_a \bar{t}_i^a t_{kl}^{ca}$ |
| $\bar{\gamma}_{ij}^{ab} \leftarrow \frac{1}{2}t_i^a t_j^b + \frac{1}{4}t_{ij}^{ab}$ | $\bar{\gamma}_{jk}^{bc} \leftarrow -\sum_{ai} \bar{t}_i^a t_k^c t_j^a t_i^b$ | $\bar{\gamma}_{kl}^{cd} \leftarrow -\frac{1}{2}\sum_{ai} \bar{t}_i^a(t_l^a t_{ki}^{cd} + t_i^d t_{ij}^{ab})$ |
| $\bar{\gamma}_{ak}^{bc} \leftarrow \sum_i \bar{t}_i^a t_k^c t_i^b$ | $\bar{\gamma}_{ak}^{ik} \leftarrow \bar{t}_i^a$ | $\bar{\gamma}_{rs}^{pq} \leftarrow \frac{1}{4}\sum_{aibj} \bar{t}_{ij}^{ab} \Lambda_{ap}^p \Lambda_{bq}^p \Lambda_{ir}^h \Lambda_{js}^h$ |
| $\bar{\gamma}_{jk}^{ik} \leftarrow -\sum_a \bar{t}_i^a t_j^a$ | $\bar{\gamma}_{ak}^{ic} \leftarrow \bar{t}_i^a t_k^c$ | $\bar{\gamma}_{ak}^{ck} \leftarrow \frac{1}{2}\sum_{bij} \bar{t}_{ij}^{ab} t_{ij}^{cb}$ |
| $\bar{\gamma}_{jk}^{ic} \leftarrow -\sum_a \bar{t}_i^a t_k^c t_j^a$ | $\bar{\gamma}_{ak}^{bk} \leftarrow \sum_i \bar{t}_i^a t_i^b$ | $\bar{\gamma}_{kl}^{il} \leftarrow -\frac{1}{2}\sum_{abj} \bar{t}_{ij}^{ab} t_{kj}^{ab}$ |
| $\bar{\gamma}_{jk}^{bk} \leftarrow -\sum_{ai} \bar{t}_i^a t_j^a t_i^b$ | | |

**DATA AVAILABILITY**

The data that support the findings of this study are available within the article.

**REFERENCES**

[1] O. Christiansen, H. Koch, and P. Jørgensen, Chem. Phys. Lett. **243**, 409 (1995).

[2] A. Hellweg, S. Grün, and C. Hättig, Phys. Chem. Chem. Phys. **10**, 4119 (2008).

[3] C. Hättig and F. Weigend, J. Chem. Phys. **113**, 5154 (2000).

[4] C. Hättig and K. Hald, Phys. Chem. Chem. Phys. **4**, 2111 (2002).

[5] H. H. Falden, K. R. Falster-Hansen, K. L. Bak, S. Rettrup and S. P. Sauer, J. Chem. Phys. A **113**, 11995 (2009).

[6] A. Tajti and P. G. Szalay, J. Chem. Theory Comput. **15**, 5523 (2019).

[7] S. Haldar, T. Mukhopadhyay, and A. K. Dutta, J. Chem. Phys. **156**, 014110 (2022).

[8] C. Hättig, J. Chem. Phys. **118**, 7751 (2003).

[9] A. Köhn and C. Hättig, J. Chem. Phys. **119**, 5021 (2003).

[10] N. O. Winter and C. Hättig, Chem. Phys. **401**, 217 (2012).

[11] K. Ledermüller and M. Schütz, J. Chem. Phys. **140**, 164113 (2014).

[12] A. Tajti, J. F. Stanton, D. A. Matthews, S. Rettrup and P. G. Szalay, J. Chem. Theory Comput. **14**, 5859 (2018).

[13] O. Christiansen, A. Halkier, H. Koch, P. Jørgensen, and T. Helgaker, J. Chem. Phys. **108**, 2801 (1998).

[14] C. Hättig, O. Christiansen, and P. Jørgensen, J. Chem. Phys. **108**, 8331 (1998).

[15] C. Hättig and A. Köhn, J. Chem. Phys. **117**, 6939 (2002).

[16] M. Schütz, J. Chem. Phys. **142**, 214103 (2015).

[17] P. Baudin, T. Kjærgaard, and K. Kristensen, J. Chem. Phys. **146**, 144107 (2017).

[18] O. Christiansen, J. F. Stanton, and J. Gauss, J. Chem. Phys. **108**, 3987 (1998).

[19] D. H. Friese, N. O. Winter, P. Balzerowski, R. Schwan, and C. Hättig, J. Chem. Phys. **136**, 174106 (2012).

[20] C. Hättig, A. Köhn, and K. Hald, J. Chem. Phys. **116**, 5401 (2002).

[21] O. Vahtras, J. Almlöf, and M. W. Feyereisen, Chem. Phys. Lett. **213**, 514 (1993).

[22] H. J. Werner, F. R. Manby, and P. J. Knowles, J. Chem. Phys. **118**, 8149 (2003).

[23] T. B. Pedersen, A. M. Sánchez de Merás, and H. Koch, J. Chem. Phys. **120**, 8887 (2004).

[24] F. Aquilante, R. Lindh, and T. Bondo Pedersen, J. Chem. Phys. **127**, 114107 (2007).

[25] P. Baudin, J. S. Marín, I. G. Cuesta, and A. M. Sánchez de Merás, J. Chem. Phys. **140**, 104111 (2014).

[26] M. Feyereisen, G. Fitzgerald, and A. Komornicki, Chem. Phys. Lett. **208**, 359 (1993).

[27] K. Eichkorn, O. Treutler, H. Öhm, M. Häser, and R. Ahlrichs, Chem. Phys. Lett. **240**, 283 (1995).

[28] D. E. Bernholdt and R. J. Harrison, Chem. Phys. Lett. **250**, 477 (1996).

[29] R. Baer, D. Neuhauser, and E. Rabani, Phys. Rev. Lett. **111**, 106402 (2013).

[30] D. Neuhauser, E. Rabani, Y. Cytter, and R. Baer, J. Phys. Chem. A **120**, 3071 (2016).

[31] D. Neuhauser, E. Rabani, and R. Baer, J. Chem. Theory Comput. **9**, 24 (2013).

[32] Q. Ge, Y. Gao, R. Baer, E. Rabani, and D. Neuhauser, J. Phys. Chem. Lett. **5**, 185 (2014).

[33] T. Y. Takeshita, W. A. de Jong, D. Neuhauser, R. Baer, and E. Rabani, J. Chem. Theory Comput. **13**, 4605 (2017).

[34] D. Neuhauser, R. Baer, and D. Zgid, J. Chem. Theory Comput. **13**, 5396 (2017).

[35] T. Y. Takeshita, W. Dou, D. G. A. Smith, W. A. de Jong, R. Baer, D. Neuhauser, and E. Rabani, J. Chem. Phys. **151**, 044114 (2019).

[36] W. Dou, T. Y. Takeshita, M. Chen, R. Baer, D. Neuhauser, and E. Rabani, J. Chem. Theory Comput. **15**, 6703 (2019).

[37] W. Dou, M. Chen, T. Y. Takeshita, R. Baer, D. Neuhauser, and E. Rabani, J. Chem. Phys. **153**, 074113 (2020).

[38] L. Mejía, S. Sharma, R. Baer, G. K. L. Chan, and E. Rabani, J. Chem. Theory Comput. **20**, 7494 (2024).

[39] E. Rabani, R. Baer, and D. Neuhauser, Phys. Rev. B **91**, 235302 (2015).

[40] J. Lee and D. R. Reichman, J. Chem. Phys. **153**, 044131 (2020).

[41] C. Zhao, J. Lee, and W. Dou, J. Chem. Phys. A **128**, 9302 (2024).

[42] C. Zhao, Q. Ou, J. Lee, and W. Dou, J. Chem. Theory Comput. **20**, 5188 (2024).

[43] J. Almlöf, Chem. Phys. Lett. **181**, 319 (1991).

[44] M. Häser and J. Almlöf, J. Chem. Phys. **96**, 489 (1992).

[45] M. Häser, Theor. Chim. Acta. **87**, 147 (1993).

[46] Y. Jung, R. C. Lochan, A. D. Dutoi, and M. Head-Gordon, J. Chem. Phys. **121**, 9793 (2004).

[47] D. Kats and M. Schütz, J. Chem. Phys. **131**, 124117 (2009).

[48] K. Freundorfer, D. Kats, T. Korona, and M. Schütz, J. Chem. Phys. **133**, 244110 (2010).

[49] N. O. Winter and C. Hättig, J. Chem. Phys. **134**, 184101 (2011).

[50] K. Ledermüller, D. Kats, and M. Schütz, J. Chem. Phys. **139**, 084111 (2013).

[51] B. Helmich-Paris and L. Visscher, J. Comput. Phys. **321**, 927 (2016).

[52] F. Sacchetta, D. Graf, H. Laqua, M. A. Ambroise, J. Kussmann, A. Dreuw, and C. Ochsenfeld, J. Chem. Phys. **157**, 104104 (2022).

[53] M. A. Ambroise, F. Sacchetta, D. Graf, C. Ochsenfeld, and A. Dreuw, J. Chem. Phys. **158**, 124121 (2023).

[54] D. G. A. Smith, L. A. Burns, A. C. Simmonett, R. M. Parrish, M. C. Schieber, R. Galvelis, P. Kraus, H. Kruse, R. Di Remigio, A. Alenaizan *et al.*, J. Chem. Phys. **152**, 184108 (2020).

[55] Q. Sun, X. Zhang, S. Banerjee, P. Bao, M. Barbry, N. S. Blunt, N. A. Bogdanov, G. H. Booth, J. Chen, Z.-H. Cui *et al.*, J. Chem. Phys. **153**, 024109 (2020).

[56] Y. Shao, Z. Gan, E. Epifanovsky, A. T. B. Gilbert, M. Wormit, J. Kussmann, A. W. Lange, A. Behn, J. Deng, X. Feng *et al.*, Mol. Phys. **113**, 184 (2015).

[57] H. Koch and P. Jørgensen, J. Chem. Phys. **93**, 3333 (1990).

[58] H. Koch, R. Kobayashi, A. Sanchez de Merás, and P. Jørgensen, J. Chem. Phys. **100**, 4393 (1994).

[59] S. V. Levchenko, T. Wang, and A. I. Krylov, J. Chem. Phys. **122**, 224106 (2005).

[60] M. Schreiber, M. R. Silva-Junior, S. Sauer, and W. Thiel, J. Chem. Phys. **128**, 134110 (2008).

[61] T. Helgaker, J. Gauss, P. Jørgensen, and J. Olsen, J. Chem. Phys. **106**, 6430 (1997).

[62] M. J. van Setten, F. Caruso, S. Sharifzadeh, X. Ren, M. Scheffler, F. Liu, J. Lischner, L. Lin, J. R. Deslippe *et al.*, J. Chem. Theory Comput. **11**, 5665 (2015).

[63] F. J. Lovas and D. R. Johnson, J. Chem. Phys. **55**, 41 (1971).

[64] R. B. Hake and K. E. Banyard, J. Chem. Phys. **43**, 657 (1965).

[65] F. Pawłowski, P. Jørgensen, J. Olsen, F. Hegelund, T. Helgaker, J. Gauss, K. L. Bak, and J. F. Stanton, J. Chem. Phys. **116**, 6482 (2002).

[66] S. G. Balasubramani, G. P. Chen, S. Coriani, M. Diedenhofen, M. S. Frank, Y. J. Franzke, F. Furche, R. Grotjahn, M. E. Harding, and C. Hättig *et al.*, J. Chem. Phys. **152**, 184107 (2020).